\documentclass{IEEEtran} 
\def\BibTeX{{\rm B\kern-.05em{\sc i\kern-.025em b}\kern-.08em
     T\kern-.1667em\lower.7ex\hbox{E}\kern-.125emX}}

\usepackage{epsf,cite,amsmath,amscd,amssymb,graphicx,latexsym,multicol}
\usepackage{algorithm,algorithmic}

\usepackage{comment}

\newtheorem{proposition}{Proposition}

\newtheorem{lemma}{Lemma}

\centerfigcaptionstrue

\begin{document}

\bibliographystyle{ieeetr}

\title{{{A Game-Theoretic Approach to Energy-Efficient Modulation in CDMA Networks with Delay QoS Constraints}\vspace{0.5cm}}\thanks{This research was supported by the National
Science Foundation under Grant ANI-03-38807, the US Army under
MURI award W911NF-05-1-0246, and the ONR under award
N00014-05-1-0168. Parts of this work have been presented at the
IEEE Radio and Wireless Symposium (RWS), Long Beach, CA, in
January 2007.}}
\author{\normalsize{Farhad Meshkati,~\IEEEmembership{Member,~IEEE,} Andrea J. Goldsmith,~\IEEEmembership{Fellow,~IEEE,}
H. Vincent Poor,~\IEEEmembership{Fellow,~IEEE,} and Stuart C.
Schwartz,~\IEEEmembership{Life Fellow,~IEEE}} \thanks{ {{F.
Meshkati was with the Department of Electrical Engineering,
Princeton University. He is currently with Qualcomm Inc., 5775
Morehouse Dr., San Diego, CA 92121 USA (e-mail:
meshkati@qualcomm.com).}}}
\thanks{A. J. Goldsmith is with the Department of Electrical
Engineering, Stanford University, Stanford, CA 94305 USA (e-mail:
andrea@ee.stanford.edu).}
\thanks{H. V. Poor and S. C. Schwartz are with the Department of
Electrical Engineering, Princeton University, Princeton, NJ 08544
USA (e-mail: {\{poor,stuart\}@princeton.edu}).}}

\maketitle

\begin{abstract}
A game-theoretic framework is used to study the effect of
constellation size on the energy efficiency of wireless networks
for M-QAM modulation. A non-cooperative game is proposed in which
each user seeks to choose its transmit power (and possibly
transmit symbol rate) as well as the constellation size in order
to maximize its own utility while satisfying its delay
quality-of-service (QoS) constraint. The utility function used
here measures the number of reliable bits transmitted per joule of
energy consumed, and is particularly suitable for
energy-constrained networks. The best-response strategies and Nash
equilibrium solution for the proposed game are derived. It is
shown that in order to maximize its utility (in bits per joule), a
user must choose the lowest constellation size that can
accommodate the user's delay constraint. This strategy is
different from one that would maximize spectral efficiency. Using
this framework, the tradeoffs among energy efficiency, delay,
throughput and constellation size are also studied and quantified.
In addition, the effect of trellis-coded modulation on energy
efficiency is discussed.
\end{abstract}

\begin{keywords}
Energy efficiency, M-QAM modulation, trellis-coded modulation, game
theory, utility function, Nash equilibrium, delay,
quality-of-service (QoS), cross-layer design.
\end{keywords}

\section{Introduction}

Wireless networks are expected to support a variety of applications
with diverse quality-of-service (QoS) requirements. Because of the
scarcity of network resources (i.e., energy and bandwidth), radio
resource management is crucial to the performance of wireless
networks. The goal is to use the network resources as efficiently as
possible while providing the required QoS to the users. Adaptive
modulation has been shown to be an effective method for improving
the spectral efficiency in wireless networks (see for example
\cite{WebbSteele,GoldsmithChua97,GoldsmithChua98,Yoo05}). However,
the focus of many of the studies to date has been on maximizing the
throughput of the network, and the impact of the modulation order on
energy efficiency has not been studied to the same extent. Recently,
the authors of \cite{Cui05} have studied modulation optimization for
an energy-constrained time-division-multiple-access (TDMA) network.
For such a network, they have used a convex-optimization approach to
obtain the best modulation strategy that minimizes the total energy
consumption under throughput and delay constraints.

Game-theoretic approaches to power control have recently attracted
considerable attention (see, for example, \cite{MackenzieWicker01,
GoodmanMandayam00, Saraydar02, Xiao01, Zhou01, Alpcan, Sung,
MeshkatiTCOM, MeshkatiJSAC, MeshkatiISIT, MeshkatiDelay}). In
\cite{MackenzieWicker01}, the authors provide motivations for
using game theory to study communication systems, and in
particular power control. In \cite{GoodmanMandayam00}, power
control is modeled as a non-cooperative game in which users choose
their transmit powers in order to maximize their utilities, where
utility is defined as the ratio of throughput to transmit power. A
game-theoretic approach to joint power control and receiver design
is presented in \cite{MeshkatiTCOM}, and power control for
multicarrier systems is studied in  \cite{MeshkatiJSAC}. The
authors in \cite{Saraydar02} use pricing to obtain a more
efficient solution for the power control game. Similar approaches
are taken in \cite{Xiao01,Zhou01,Alpcan,Sung} for different
utility functions. Game-theoretic approaches to power control in
delay-constrained networks are proposed in \cite{MeshkatiISIT,
MeshkatiDelay}.

In this work, we use a game-theoretic approach to study the effects
of modulation on energy efficiency of code-division-multiple-access
(CDMA) networks in a \emph{competitive} multiuser setting. Focusing
on M-QAM modulation, we propose a \emph{non-cooperative} game in
which each user chooses its strategy, which includes the choice of
the transmit power, transmit symbol rate and constellation size, in
order to maximize its own utility while satisfying its QoS
constraints. The utility function used here measures the number of
reliable bits transmitted per joule of energy consumed, and is
particularly suitable for energy-constrained networks. We derive the
best-response strategies and Nash equilibrium solution for the
proposed game. In addition, using our non-cooperative game-theoretic
framework, we quantify the tradeoffs among energy efficiency, delay,
throughput and modulation order. The effect of coding on energy
efficiency is also studied and quantified using the proposed
game-theoretic approach. In addition, our framework allows us to
illustrate the tradeoff between energy efficiency and spectral
efficiency.

The remainder of this paper is organized as follows. The system
model and definition of the utility function are given in
Section~\ref{systemmodel}. We then present a power control game with
no delay constraints in Section~\ref{pcg} and derive the
corresponding Nash equilibrium solution. A delay-constrained power
control game is proposed in Section~\ref{pcgd} and the corresponding
best-response strategies and Nash equilibrium solution are derived.
The analysis is extended to coded systems in Section~\ref{pcgtcm}.
Numerical results and conclusions are given in
Sections~\ref{numericalresults}~and~\ref{conclusion}, respectively.

\section{System Model} \label{systemmodel}

We consider a direct-sequence CDMA (DS-CDMA) wireless network in
which the users' terminals are transmitting to a common
concentration point (e.g., a cellular base station or an access
point). The system bandwidth is assumed to be $B$ Hz. Let $R_{s,k}$
and $p_k$ be the symbol rate and the transmit power for user $k$,
respectively. In this work, we focus on M-QAM modulation. Hence,
each symbol is assumed to be complex to represent the in-phase and
quadrature components.  For the M-QAM modulation, the number of bits
transmitted per symbol is given by $$b=\log_2 M.$$ Since there is a
one-to-one mapping between $M$ and $b$, we sometimes refer to $b$ as
the constellation size. We focus on \emph{square} M-QAM modulation,
i.e., $M\in\{4,16,64,\cdots\}$ or equivalently
$b\in\{2,4,6,\cdots\}$, since there are exact expressions for the
symbol error probability of square M-QAM modulation (see
\cite{GoldsmithBook}). We can easily generalize our analysis to
include odd values of $b$ by using an approximate expression for the
symbol error probability.

We define the utility function of a user as the ratio of its
throughput to its transmit power, i.e.,
\begin{equation}\label{eq2}
   u_k = \frac{T_k}{p_k} \ .
\end{equation}
This utility function is similar to the one used in
\cite{GoodmanMandayam00} and \cite{Saraydar02}. Throughput in
\eqref{eq2} is defined as the net number of information bits that
are transmitted without error per unit time (it is sometimes
referred to as \emph{goodput}), and is expressed as
\begin{equation}\label{eq3}
   T_k = R_k f(\gamma_k)
\end{equation}
where $R_k= b_k R_{s,k}$ is the transmission rate, $\gamma_k$ is the
output signal-to-interference-plus-noise ratio (SIR) for user $k$,
and $f(\gamma_k)$ is the ``efficiency function" which represents the
packet success rate (PSR) for the $k$th user. We require that
$f(0)=0$ to ensure that $u_k=0$ when $p_k=0$. In general, the
efficiency function depends on the modulation, coding and packet
size. We assume an automatic-repeat-request (ARQ) mechanism in which
the user keeps retransmitting a packet until the packet is received
at the access point without any errors. Based on (\ref{eq2}) and
(\ref{eq3}), the utility function for user $k$ can be written as
\begin{equation}\label{eq4}
   u_k = R_k \frac{f(\gamma_k)}{p_k}\ .
\end{equation}
This utility function, which has units of \emph{bits/joule},
measures the number of reliable bits that are transmitted per joule
of energy consumed, and is particularly suitable for
energy-constrained networks.

Let us for now focus on a specific user and drop the subscript $k$.
Assuming a packet size of $L$ bits, the packet success rate for
square M-QAM modulation is given by
\begin{equation}\label{eq15}
P_{\textrm{success}}(b,\gamma)= \left( 1- \alpha_b Q(\sqrt{\beta_b
\gamma})\right)^{\frac{2L}{b}}
\end{equation}
where
\begin{equation}\label{eq15a}
    \alpha_b = 2\left(1-2^{-b/2}\right)\ ,
\end{equation}
and
\begin{equation}\label{eq15b}
    \beta_b= \frac{3}{2^b-1} \ .
\end{equation}
Here, $\gamma$ represents the symbol SIR and $Q(\cdot)$ is the
complementary cumulative distribution function of the standard
Gaussian random variable. Note that at $\gamma=0$, we have
$P_{\textrm{success}}~=~2^{-L}~\neq~0$. Since we require the
efficiency function to be zero at zero transmit power, we define
\begin{equation}\label{eq16}
f_b(\gamma)= \left( 1- \alpha_b Q(\sqrt{\beta_b \gamma})
\right)^{\frac{2L}{b}} - 2^{-L}.
\end{equation}
Note that $2^{-L}\simeq 0$ when $L$ is large (e.g., $L=100$).

A non-cooperative power control game, in general, can be expressed
as ${G}=[{\mathcal{K}}, \{\mathcal{A}_k \}, \{{u}_k \}]$ where
${\mathcal{K}}=\{1, ... , K \}$ is the set of users/players,
$\mathcal{A}_k$ is the strategy set for the $k${th} user, and $u_k$
is the utility function given by \eqref{eq4}. Each user decides what
strategy to choose from its strategy set in order to maximize its
own utility. Hence, the best-response (i.e. utility-maximizing)
strategy of user $k$ is given by the solution of
\begin{equation}\label{eq13}
   \max_{a_k \in \mathcal{A}_k } \ u_k  \ \ \ \ \equiv \ \  \max_{a_k \in \mathcal{A}_k
   } R_k \frac{f(\gamma_k)}{p_k}\ .
\end{equation}
For this game, a \emph{Nash equilibrium} is a set of strategies
$(a_1^*,\cdots, a_K^*)$ such that no user can unilaterally improve
its own utility \cite{FudenbergTiroleBook91}, that is,
\begin{equation}\label{Ch1/eq2}
    u_k(a_k^*, a_{-k}^*)\geq u_k(a_k,a_{-k}^*) , \ \textrm{for
    all} \ a_k\in \mathcal{A}_k \ \textrm{and} \ \ k=1,\cdots,K .
\end{equation}

In this work, we propose non-cooperative games in which the actions
open to each user are the choice of transmit power, and possibly
transmit symbol rate, as well as the choice of \emph{constellation
size}.

\section{Power Control Game with M-QAM Modulation}\label{pcg}

Consider a DS-CDMA network with $K$ users and express the
transmission rate of user $k$ as
\begin{equation}\label{eq14}
R_k= b_k R_{s,k}
\end{equation}
where $b_k$ is the number of information bits per symbol and
$R_{s,k}$ is the symbol rate. Let us for now assume that users have
no delay constraints. We propose a power control game in which each
user seeks to choose its constellation size and transmit power in
order to maximize its own utility, i.e.,
\begin{equation}\label{eq14b}
    \max_{b_k , p_k} R_k \frac{f_{b_k}(\gamma_k)}{p_k}\ \ \ \textrm{for} \
    \ k=1,\cdots,K ,
\end{equation}
where $b_k \in \{2,4,6,\cdots\}$ and $p_k \in [0, P_{max}]$ with
$P_{max}$ being the maximum allowed transmit power. Throughout this
work, we assume $P_{max}$ is large.

For all linear receivers, the output SIR for user $k$ can be written
as
\begin{equation}\label{eq17}
    \gamma_k= (B/R_{s,k}) p_k\ \hat{h}_k
\end{equation}
where $B$ is the system bandwidth and $\hat{h}_k$ is the effective
channel gain which depends on the channel gain of user $k$ and on
the channel gains and transmit powers of other users in the network
but is independent of the transmit power and rate of user $k$. For
example, for a matched-filter receiver, $\hat{h}_k$ is given by
$$\hat{h}_k= \frac{h_k}{\sigma^2 + \sum_{j\neq k} p_j h_j} ,$$ where $h_k$ is the channel gain
for user $k$, and $\sigma^2$ is the noise power. Therefore, the
utility function in \eqref{eq4} can be written as
\begin{equation}\label{eq17b}
    u_k = B \hat{h}_k\ b_k
    \frac{f_{b_k}(\gamma_k)}{\gamma_k} .
\end{equation}
Based on \eqref{eq17b}, and by dropping the subscript $k$ for
convenience, the maximization in \eqref{eq14b} can be written as
\begin{equation}\label{eq18}
    \max_{b , \gamma} B \hat{h}\ b
    \frac{f_{b}(\gamma)}{\gamma} .
\end{equation}
Since for a given user, $B$ and $\hat{h}$ are fixed, maximizing the
user's utility is equivalent to maximizing $b f_b(\gamma)/\gamma$
for that user. It is important to observe that, for a given $b$,
specifying the operating SIR completely specifies the utility
function. Let us for now fix the symbol rate $R_s$ and the
constellation size. Taking the derivative of \eqref{eq18} with
respect to $\gamma$ and equating it to zero, we conclude that the
utility of a user is maximized when its output SIR, $\gamma$, is
equal to $\gamma_b^*$, which is the (positive) solution of
\begin{equation}\label{eq19}
    f_b(\gamma) = \gamma f'_b(\gamma).
\end{equation}
It is shown in \cite{Rod03b} that for an S-shaped\footnote{An
increasing function is S-shaped if there is a point above which
the function is concave, and below which the function is convex.}
(sigmoidal) efficiency function, ${f(\gamma_k)=\gamma_k \
f'(\gamma_k)}$ has a unique solution. It can easily be verified
that $f_b(\gamma)$ given by \eqref{eq16} is S-shaped. Note that
$\gamma_b^*$ is (uniquely) determined by physical-layer parameters
such as packet size, modulation, and coding.

\begin{figure}
\begin{center}
 \leavevmode \hbox{\epsfysize=3in \epsfxsize=3.5in
\epsffile{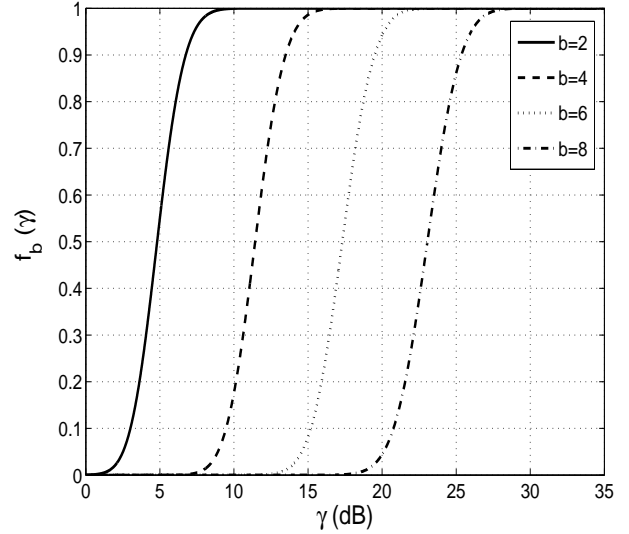}}
\end{center} \caption{Efficiency function (packet success rate) as a function of SIR for different constellation sizes.}\label{fig1}
\end{figure}

\begin{figure}
\begin{center}
 \leavevmode \hbox{\epsfysize=3in \epsfxsize=3.5in
\epsffile{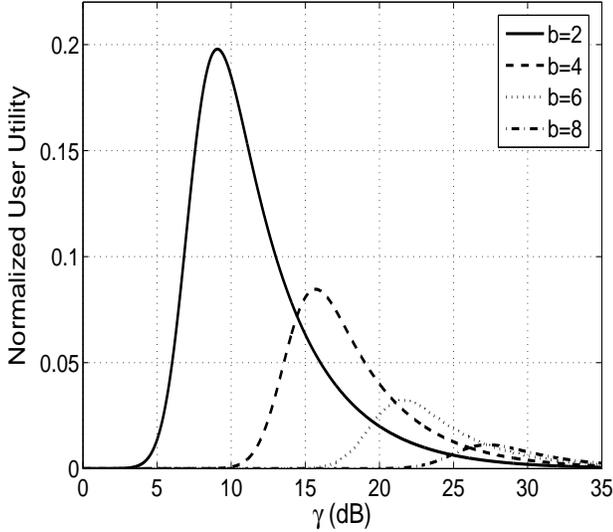}}
\end{center} \caption{Normalized user utility as a function of SIR for different constellation sizes.}\label{fig2}
\end{figure}

Assuming that $\gamma_b^*$ is feasible, the maximum utility is
hence given by
\begin{equation}\label{eq20}
    u_b^*=B \hat{h}\ b
    \frac{f_{b}(\gamma_b^*)}{\gamma_b^*}\ .
\end{equation}
Based on \eqref{eq16}, it can be shown that $\gamma_b^*$ is
(approximately) given by the solution of
$$\frac{\alpha_b L}{b} \sqrt{\frac{\beta_b \gamma}{2\pi}} \
e^{-\frac{\beta_b \gamma}{2}} + \alpha_b Q(\sqrt{\beta_b \gamma})
= 1.$$ We can compute $\gamma_b^*$ numerically for different
values of $b$. Table~\ref{table1} summarizes the results for a
system with $L=100$ bits (i.e., 100 bits per packet).
\begin{table}
\begin{center} \caption{Summary of the parameters corresponding to the utility-maximizing solutions for
different modulation orders for an uncoded system with 100 bits
per packet}\label{table1}
\begin{tabular}{|c|c|c|c|c|c|c|c|c|}
  \hline
  $b$ & $\alpha_b$ & $\beta_b$ & $\gamma_b^*$(dB) & $f_b(\gamma_b^*)$ & $b/\gamma_b^*$(dB) & $b f_b(\gamma_b^*)/ \gamma_b^*$ \\
  \hline
  \hline
  2 & 1 & 1 &  9.1 & 0.801 & -6.1 & 0.1978\\
  4 & 1.5 & 0.2 &  15.7 & 0.785 & -9.7 & 0.0846\\
  6 & 1.75 & 0.0476 &  21.6 & 0.771 & -13.8 & 0.0322\\
  8 & 1.875 & 0.0118 &  27.3 & 0.757 & -18.3 & 0.0112\\
  10 &1.9375 & 0.0029 &  33.0 & 0.743 &-23.0 & 0.0037\\

  \hline
\end{tabular}
\end{center}
\end{table}
It is observed from Table~\ref{table1} that the user's utility is
maximized when $b=2$ (i.e., QPSK modulation). This is because, as
$b$ increases, the linear increase in the throughput is dominated
by the exponential increase in the required transmit power (which
results from the exponential increase in $\gamma_b^*$). Therefore,
it is best for a user to use QPSK modulation.\footnote{BPSK and
QPSK are equivalent in terms of energy efficiency, but QPSK has a
higher spectral efficiency.} Figs.~\ref{fig1}~and~\ref{fig2} show
the efficiency function and the normalized user utility (i.e.,
$\frac{u_b}{B \hat{h}}$) as a function of the SIR for different
choices of $b$.

It can be shown that if $\gamma_b^*$ is not feasible, the user's
utility is maximized when the transmit power is equal to $P_{max}$
(see \cite{Saraydar02} and \cite{MeshkatiTCOM}). Now, if we assume
that $P_{max}$ is large, then for a matched filter receiver, in
order for users $1,\cdots,K$ to achieve output SIRs equal to
$\gamma_1$,$\cdots$,$\gamma_K$, respectively, the transmit powers
must be equal to
\begin{equation} \label{eq20d}
p_k=\frac{\sigma^2}{h_k}\left(\frac{\Phi_k}{1-\sum_{j=1}^K
\Phi_j}\right) ,
\end{equation}
where
\begin{equation}\label{eq20e}
\Phi_k=\left(1+\frac{B}{R_{s,k}\gamma_{k}}\right)^{-1} ,
\end{equation}
for $k=0,\cdots,K$. Therefore, $\gamma_1$,$\cdots$,$\gamma_K$ are
feasible if and only if $$\sum_{k=1}^K
\left(1+\frac{B}{R_{s,k}\gamma_{k}}\right)^{-1} < 1.$$

So far, we have shown that at Nash equilibrium (if it exists),
QPSK modulation must be used by each user (otherwise a user can
always improve its utility by switching to QPSK and reducing its
transmit power), and each user's transmit power is chosen to
achieve the $\gamma^*$ corresponding to the QPSK modulation at the
output of the receiver (i.e., 9.1dB according to
Table~\ref{table1}). The existence of the Nash equilibrium for the
proposed game can be shown via the quasiconcavity of each user's
utility function in its own power \cite{FudenbergTiroleBook91}.
Furthermore, because of the uniqueness of $\gamma^*$ and the
one-to-one correspondence between the transmit power and the
output SIR (see \eqref{eq20d}), this equilibrium is
unique.\footnote{Please note that throughout this paper, SIR
refers to the output signal-to-interference-plus-noise ratio.
Based on \eqref{eq20d} and \eqref{eq20e}, specifying the SIRs of
the users uniquely determines their transmit powers and vice
versa.}

In addition, we observe that the energy-efficient strategy is not
spectrally-efficient. Incorporating the choice of the modulation
order into our utility maximization allows us to trade off
\emph{energy efficiency} with \emph{spectral efficiency}. For the
same bandwidth and symbol rate, as a user switches to a
higher-order modulation, the spectral efficiency for the user
improves but its energy efficiency degrades.

\section{Delay-Constrained Power and Rate Control Game with M-QAM
Modulation}\label{pcgd}

In Section~\ref{pcg}, we showed that for our utility function, it is
best for a user to use the lowest-order modulation. We now extend
the analysis to the case in which the users have delay QoS
requirements. Our goal in this part is to study the effects of
constellation size on energy efficiency and delay. We consider a
game in which each user seeks to choose its transmit power, symbol
rate and constellation size to maximize its own utility while
satisfying its delay QoS constraint. The delay QoS constraint
considered here is in terms of the average delay and includes both
\emph{transmission} and \emph{queuing} delays. More discussion on
the delay performance can be found in \cite{MeshkatiDelay}. It
should be noted that an average-delay constraint may not be
sufficient for applications with hard delay requirements (see
\cite{Holliday02}).

\subsection{Delay Model}

Let us assume that the incoming packets for user $k$ have a Poisson
distribution with parameter $\lambda_k$ which represents the average
packet arrival rate with each packet consisting of $L$ bits. The
source rate (in bits per second) is hence given by $L\lambda_k$.
The user transmits the arriving packets at a rate $R_k= b_k R_{s,k}$
(bps) and with a transmit power equal to $p_k$ Watts. We assume an
ARQ mechanism for packet transmission. Also, the incoming packets
are assumed to be stored in a queue and transmitted in a
first-in-first-out (FIFO) fashion. The packet success probability
(per transmission) as before is represented by the efficiency
function $f_b(\gamma)$.

\begin{figure}
\centering
\includegraphics[width=3in]{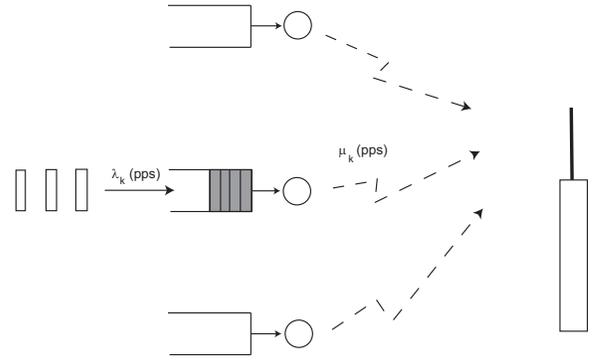}
\caption{System model based on an M/G/1 queue.} \label{fig3}
\end{figure}

Focusing on a specific user and dropping the subscript $k$, we can
represent the combination of the user's queue and wireless link as
an M/G/1 queue (as shown in Fig.~\ref{fig3}) where the service
time, $S$, has the following probability mass function (PMF):
\begin{equation}\label{eq23}
    \textrm{Pr}\{S=m\tau\}= f_b(\gamma)
    \left(1-f_b(\gamma)\right)^{m-1}  \  \textrm{for} \ m=1, 2,
    \cdots
\end{equation}
with $\tau$ being the packet transmission time which is given by
\begin{equation}\label{eq22}
    \tau = \frac{L}{b R_s} + \epsilon \simeq \frac{L}{b R_s} .
\end{equation}
Here, $\epsilon$ represents the time taken for the user to receive
an ACK/NACK from the access point. We assume $\epsilon$ is
negligible compared to $\frac{L}{b R_s}$. Based on \eqref{eq23}, the
service rate, $\mu$, is given by
\begin{equation}\label{eq24}
    \mu=\frac{1}{\mathbb{E}\{S\}}= \frac{f_b(\gamma)}{\tau}=
    R_s \frac{b f_b(\gamma)}{L} ,
\end{equation}
and the load factor $\rho=\frac{\lambda}{\mu}=\frac{\lambda
\tau}{f_b(\gamma)}$. Therefore, the average service rate is affected
by the constellation size through a linear factor $b$ as well as the
efficiency function $f_b(\gamma)$. To keep the queue stable, we must
have $\rho<1$ or $f_b(\gamma)>\lambda \tau$.

Now, let $W$ be a random variable representing the total packet
delay for the user. The delay includes both transmission and
queuing delays. Using the Pollaczek-Khintchine formula for the
M/G/1 queue considered here, the average packet delay is given by
(see \cite{GrossBook85})
\begin{equation}\label{eq25}
    \bar{W} = \tau \left(\frac{1-\frac{\lambda
    \tau}{2}}{f_b(\gamma)-\lambda \tau}\right) \ \ \
    \textrm{with} \ f_b(\gamma)>\lambda \tau .
\end{equation}

We specify the delay QoS constraint of a user by an upper bound on
the average packet delay, i.e., we require
\begin{equation}\label{eq26}
    \bar{W} \leq D .
\end{equation}
This delay constraint can equivalently be expressed as
\begin{equation}\label{eq27}
    f_b(\gamma) \geq \eta_b
\end{equation}
where
\begin{equation}\label{eq28}
    \eta_b = \frac{L \lambda}{b R_s}  + \frac{L}{b R_s D} -
    \frac{L^2 \lambda }{2 b^2 R_s^2 D} .
\end{equation}

Note that \eqref{eq27} is equivalent to the condition $$\gamma\geq
\hat{\gamma_b}$$ where
\begin{equation}\label{eq29}
    \hat{\gamma}_b=f_b^{-1}(\eta_b) \ .
\end{equation}
Based on \eqref{eq16}, $\hat{\gamma}_b$ is given by
\begin{equation}\label{eq29b}
    \hat{\gamma}_b\simeq \frac{1}{\beta_b} \left[Q^{-1}
    \left(\frac{1-\eta^{b/2L}}{\alpha_b}\right)\right]^2 .
\end{equation}
This means that the delay constraint in \eqref{eq26} translates
into a lower bound on the output SIR. It should be noted that
since the upper bound on the average delay must be at least as
large as the transmission time, i.e., ${D}\geq{\tau}$, we must
have that $bR_s\geq L/D$. This automatically implies that
$\eta_b>0$. Also, since $0\leq f_b(\gamma) \leq 1$, \eqref{eq27}
is possible only if $\eta_b <1$.\footnote{Note that $f(\gamma)=1$
requires an infinite SIR which is not practical.}

\subsection{The Proposed Game}\label{the game}

We propose a game in which each user chooses its transmit power and
symbol rate as well as its constellation size in order to maximize
its own utility while satisfying its delay requirement. Fixing the
other users' transmit powers and rates, the best-response strategy
for the user of interest is given by the solution of the following
constrained maximization:\vspace{-0cm}
\begin{equation}\label{eq30}
    \max_{p, R_s, b} \ u \ \ \ \textrm{s.t.} \ \ \ \bar{W} \leq D \ ,
\end{equation}
or equivalently
\begin{equation}\label{eq31}
    \max_{\gamma, R_s, b} \ b \frac{f_b(\gamma)}{\gamma} \ \ \
     \textrm{s.t.} \ \  \gamma \geq \hat{\gamma}_b \ \ \textrm{and} \ \ 0\leq \eta_b<1 \
     .
\end{equation}
\begin{proposition} \label{prop1}
For a fixed $b$, the source rate $\lambda$ and the delay constraint
$D$ are feasible if and only if
\begin{equation}\label{eq31a}
    \frac{L \lambda}{b B}  + \frac{L}{b B D} -
    \frac{L^2 \lambda }{2 b^2 B^2 D} < 1 \ ,
\end{equation}
where $B$ is the system bandwidth.
\end{proposition}
\begin{proof}
For $\lambda$ and $D$ to be feasible, we must have $\eta_b<1$
where $\eta_b$ is given by \eqref{eq28}. Also, since $\eta_b$ is a
decreasing function of $R_s$, the lowest possible value of
$\eta_b$ is achieved when $R_s=B$. Hence, it is straightforward to
see that the source rate $\lambda$ and the delay constraint $D$
are feasible if and only if \eqref{eq31a} is satisfied.
\end{proof}

Remember that $D$ cannot be smaller than $\tau$. Hence, it can be
shown that the condition $0\leq \eta_b<1$ is equivalent to $R_s
> \Omega^\infty /b$ where $$\Omega^{\infty}=\left(\frac{L}{D}\right)
\frac{1+D\lambda +\sqrt{1+ D^2 \lambda^2}}{2}.$$ Also, let us define
$\Omega_b^*$ as the rate for which $\hat{\gamma}_b=\gamma_b^*$,
i.e.,
\begin{equation} \label{eq31b}
\Omega_b^*=\left(\frac{L}{D}\right) \frac{1+D\lambda +\sqrt{1+ D^2
\lambda^2 +2(1-f_b^*)D \lambda}}{2f_b^*}
\end{equation}
where $f_b^*=f_b(\gamma_b^*)$.

\begin{proposition} \label{prop2}
For given values of $\lambda$ and $D$, the best-response strategy
for a user (i.e., the solution of \eqref{eq30}) is any combination
of $p$ and $R_s$ such that
\begin{equation}
\min\left\{\Omega_{\tilde{b}}^*/\tilde{b}, B\right\} \leq R_s \leq B
\end{equation}
and
\begin{equation}\label{eq31c}
\gamma=\left\{
         \begin{array}{ll}
           \gamma_{\tilde{b}}^*, & \hbox{if  $\Omega_{\tilde{b}}^*/\tilde{b} \leq B$;} \\
           \hat{\gamma}_{\tilde{b}}, & \hbox{if  $\Omega_{\tilde{b}}^*/\tilde{b} > B$,}
         \end{array}
       \right.
\end{equation}
where $\tilde{b}$ is the lowest constellation size for which
$\lambda$ and $D$ are feasible, $\gamma_b^*$ is the solution of
\eqref{eq19}, and $\hat{\gamma}_{b}$ is given by \eqref{eq29}.
\end{proposition}

Proof of Proposition~\ref{prop2} requires the following lemma.
\begin{lemma}\label{lemma1}
If $\eta_b> 2^{-L}$, then $\hat{\gamma}_b$ is an increasing function
of $b$ and the increase is exponential.
\end{lemma}
\begin{proof}
Let us define $x_b\equiv \frac{1-\eta^{b/2L}}{\alpha_b}$. Given
\eqref{eq15a}, we can express $x_b$ as
\begin{equation} \label{eqL1}
    x_b=\frac{1}{2}\left[\frac{1- \left( \eta_b^{1/2L} \right)^b
}{1- \left(1/\sqrt{2} \right)^b }\right] .
\end{equation}
Define $\hat{\eta}_b \equiv \eta_b^{1/2L}$. If $\eta_b> 2^{-L}$,
then $\hat{\eta}_b > 1/\sqrt{2}$. Now, if $b'>b''$, then, it is easy
to show that
$$\frac{1-(\hat{\eta}_{b''})^{b'}}{1-(1/\sqrt{2})^{b'}} >
\frac{1-(\hat{\eta}_{b''})^{b''}}{1-(1/\sqrt{2})^{b''}} . $$ Also,
since $\eta_{b'}<\eta_{b''}$, we have
$$\frac{1-(\hat{\eta}_{b'})^{b'}}{1-(1/\sqrt{2})^{b'}} > \frac{1-(\hat{\eta}_{b''})^{b'}}{1-(1/\sqrt{2})^{b'}} .$$
As a result,
$$\frac{1-(\hat{\eta}_{b'})^{b'}}{1-(1/\sqrt{2})^{b'}}>\frac{1-(\hat{\eta}_{b''})^{b''}}{1-(1/\sqrt{2})^{b''}},$$ which
implies that $x_b$ is an increasing function of $b$. According to
\eqref{eq29b}, we have $$\hat{\gamma}_b=
(1/\beta_b)\left[Q^{-1}(x_b)\right]^2 .$$ Since $x_b$ is an
increasing function of $b$, and $\beta_b$ is a decreasing function
of $b$, $\hat{\gamma}_b$ is an increasing function of $b$.
Furthermore, since $\left[Q^{-1}(x_b)\right]^2$ is increasing in $b$
and $1/\beta_b = (2^b -1)/3$ is exponentially increasing in $b$,
$\hat{\gamma}_b$ is also exponentially increasing in $b$.
\end{proof}

The $\eta_b>2^{-L}$ assumption is consistent with good design
practice. In particular, since $\eta_b$ represents the packet
success probability, $\eta_b \leq 2^{-L}$ would correspond to a
very poorly-designed system. In fact, in such a case, there would
not be any need to transmit the packets since random guessing at
the receiver would give a PSR of $2^{-L}$. We now give the proof
for Proposition~\ref{prop2}.

\begin{proof}[Proof of Proposition~\ref{prop2}]
We showed in Section~\ref{pcg} that for the unconstrained
optimization problem and for a fixed $b$, the utility is maximized
when the user's SIR is equal to $\gamma_b^*$ which is the solution
of $f_b(\gamma)=\gamma f_b'(\gamma)$. It is straightforward to show
that $\hat{\gamma}_b$ is a decreasing function of $R_s$ for all
$R_s\geq \Omega^{\infty}/b$. Therefore, for all $\Omega^{\infty}/b
\leq R_s <\Omega_b^*/b$, $\hat{\gamma}_b
> \gamma_b^*$.  This means that based on \eqref{eq17b},
a user has no incentive to transmit at a symbol rate smaller than
$\Omega_b^*/b$. Hence, for a fixed constellation size, any
combination of $p$ and $R_s \geq \Omega_b^*/b$ that results in an
output SIR equal to $\gamma_b^*$ maximizes the utility and satisfies
the delay constraint. If $\Omega_b^*/b > B$, then $\hat{\gamma}_b$
cannot be made equal to $\gamma_b^*$. In this case, the user must
transmit at the maximum symbol rate (i.e., $R_s=B$) and choose its
transmit power such that $\gamma=\hat{\gamma}_b > \gamma_b^*$ in
order to meet its delay constraint.

Now let us include the effect of constellation size. Let $b'>b''$
and consider the following cases.
\begin{itemize}
  \item If $\Omega_{b''}^*/b'' \leq B$, then we will have $\Omega_{b'}^*/b'
\leq B$. This means both $\gamma_{b'}^*$ and $\gamma_{b''}^*$ are
feasible. However, the user's utility will drop if the user moves to
a higher-order modulation. This is because the linear gain in
utility due to an increase in $b$ is dominated by the exponential
increase in the optimum operating SIR as shown in Section~\ref{pcg}.
Therefore, in this case, the user would choose the smallest $b$.
  \item If $\Omega_{b''}^*/b'' > B$ but $\lambda$ and $D$ are feasible with $b''$ (see Proposition~\ref{prop1}), then
the user's utility is maximized when the symbol rate is equal to $B$
and the SIR is equal to $\hat{\gamma}_{b''}$. On the other hand, the
user can switch to $b'>b''$. In that case, $\hat{\gamma}_{b'}$ is
smallest when the symbol rate is equal to $B$. However, based on
Lemma~\ref{lemma1}, with $R_s=B$ and $b'>b''$, we have
$\hat{\gamma}_{b'}>\hat{\gamma}_{b''}$. Furthermore, the increase in
$\hat{\gamma}_{b}$ is exponential. Since the exponential increase in
the SIR would dominate the linear increase in the rate caused by an
increase in $b$, it is best for the user to use $b''$ (i.e., the
smaller constellation size).
  \item If $\Omega_{b''}^*/b'' > B$ and $\lambda$ and $D$ are not
feasible, the user must switch to a higher constellation size and a
similar argument as above would follow.
\end{itemize}
Therefore, the user must always choose the lowest constellation size
for which the user's QoS constraint can be satisfied.
\end{proof}

Proposition~\ref{prop2} implies that, in terms of energy efficiency,
choosing the lowest-order modulation (i.e., QPSK) is the best
strategy unless the user's delay constraint is too tight. In other
words, the user would jump to a higher-order modulation only when it
is transmitting at the highest symbol rate (i.e., $R_s=B$) and still
cannot meet the delay requirement. Also, the proposition suggests
that if $\Omega_{\tilde{b}}^*/\tilde{b} < B$, the user has
infinitely many best-response strategies. In particular, the user
chooses the lowest constellation size that can accommodate the delay
constraint. Then, for that constellation, any combination of $p_k$
and $R_{s,k}$ for which $\gamma_k=\gamma_{b_k}^*$ and
$R_{s,k}\geq\Omega_{b_k}^*/b$ is a best-response strategy.

\subsection{Nash Equilibrium} \label{NE}

At Nash equilibrium, the transmit powers, symbol rates and
constellation sizes of all the users have to satisfy
Proposition~\ref{prop2} simultaneously. There are, therefore, cases
where we have infinitely many Nash equilibria. For a matched filter,
for example, the best-response transmit power of user $k$ is given
by
\begin{equation} \label{eq32}
p_k=\frac{\sigma^2}{h_k}\left(\frac{\Phi_k}{1-\sum_{j=1}^K
\Phi_j}\right) ,
\end{equation}
where
\begin{equation}\label{eq32b}
\Phi_k=\left(1+\frac{B}{R_{s,k}\gamma_{k}}\right)^{-1} ,
\end{equation}
and $\gamma_k$ and $R_{s,k}$ are determined according to
Proposition~\ref{prop2} for $k=1,\cdots,K$. We refer to $\Phi_k$
as ``size" of user $k$. $\Phi_k$ is a measure of the amount of
network resources that is consumed by the $k$th
user.\footnote{$\Phi_k$ here is a generalized version of the
definition that has been given in \cite{MeshkatiDelay}.} Note that
$R_{s,k}$'s and $\gamma_k$'s are feasible if and only if
\begin{equation}\label{eq32c}
\sum_{j=1}^K \Phi_j < 1 .
\end{equation}
Combining \eqref{eq4} with \eqref{eq32}, the utility of user $k$ at
Nash equilibrium is given by
\begin{equation}\label{eq33}
    u_k=\frac{B f(\gamma_k) h_k}{\sigma^2 \gamma_k}\left(1-
\frac{\sum_{j \neq k} \Phi_j}{1-\Phi_k}\right) \ .
\end{equation}
Therefore, the Nash equilibrium with the smallest $R_{s,k}$ achieves
the largest utility. A higher symbol rate (i.e., smaller processing
gain) for a user requires a larger transmit power by that user to
achieve the required SIR. This causes more interference for other
users in the network and forces them to raise their transmit powers
as well. As a result, the level of interference in the system
increases and the users' utilities decrease. This means that the
Nash equilibrium with
$R_{s,k}=\min\{\Omega_{\tilde{b}_k}^*/\tilde{b}_k, B\}$ and $p_k$
given by \eqref{eq32} for $k=1,\cdots,K$ is the
\emph{Pareto-dominant} Nash equilibrium.

As the delay constraint of a user becomes tighter, according to
Proposition~\ref{prop2}, the user will increase its symbol rate.
This results in an increase in the user's ``size". When the symbol
rate becomes equal to the system bandwidth, the user will increase
its SIR which again results in an increase in $\Phi$. Finally, when
the user's delay constraint is not feasible anymore, the user will
switch to a higher constellation size. This results in an
exponential increase in the required SIR, which dominates the linear
decrease in the symbol rate. Therefore, $\Phi$ increases again. This
shows that the user's ``size" increases as the delay requirement
becomes more stringent. The feasibility condition given by
\eqref{eq32c} determines the maximum number of users that can be
accommodated by the network. A tighter delay constraint results in a
larger ``size' for the user. This, in turn, results in a smaller
network capacity.

\section{Power Control Games with Trellis-Coded M-QAM Modulation}
\label{pcgtcm}

So far, we have focused on an uncoded system. In this section, we
extend our analysis to trellis-coded modulation (TCM). We consider
a trellis-coded M-QAM system in which $b$ information bits are
divided into two groups of size $n$ and $b-n$ bits, respectively.
The first group (with size $n$) is convolutionally encoded into
$\ell$ bits which are used by the coset selector to choose one of
the $2^{\ell}$ constellation subsets. The remaining $b-n$ bits are
used to choose one of the $2^{b-n}$ signal points in the selected
subset (see \cite{Ungerboeck} for more details). The code rate is
hence given by $\theta_c=n/\ell$ and the constellation size is
increased from $2^b$ to $2^{\ell+b-n}$. It is common to use a code
rate of $\theta_c=n/(n+1)$ for subset selection. For $b>2$, $n=2$
is usually a good choice.\footnote{For $b=2$, $n$ is equal to
one.}

\begin{figure}
\begin{center}
\leavevmode \hbox{\epsfysize=3.2in \epsfxsize=3.5in
\epsffile{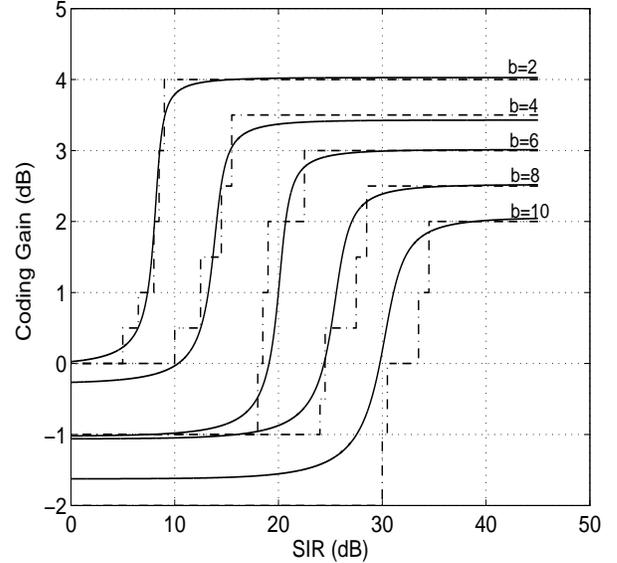}}
\end{center} \caption{Coding gain as a function of SIR for an 8-state
convolutional encoder with rate 2/3. The solid lines are the
estimates based on $\hat{G}_b$ and the dashed lines are piece-wise
linear estimates based on the data obtained from the BER curves.}
\label{fig4}
\end{figure}

For trellis-coded modulation (TCM), the efficiency function (which
represents the packet success probability) is given by
\begin{equation}\label{eq34}
f_b^{(c)}(\gamma)\simeq \left( 1- \alpha_b Q(\sqrt{\beta_b \gamma
G_b(\gamma)}) \right)^{\frac{2L}{b}} - 2^{-L} ,
\end{equation}
where $b$ is the number of \emph{information} bits per symbol and
$G_b(\cdot)$ is the effective \emph{coding gain} which in general is
a function of SIR and also depends on the modulation level. Recall
that in our proposed game, each user chooses its transmit power,
symbol rate and modulation level to maximize its own utility
function while satisfying its delay constraint. One could
potentially follow the same analysis for the coded system as the one
presented for the uncoded system by replacing $f_b(\gamma)$ with
$f_b^{(c)}(\gamma)$ given in \eqref{eq34}.  For the coded case, the
delay-constrained utility maximization can be written as
\begin{equation}\label{eq31}
    \max_{\gamma, R_s, b} \ b \frac{f_b^{(c)}(\gamma)}{\gamma} \ \ \
     \textrm{s.t.} \ \  \gamma \geq \hat{\gamma}_b \ \ \textrm{and} \ \ 0\leq \eta_b<1 \
     ,
\end{equation}
where $\hat{\gamma}_b^{(c)}=f_b^{(c) -1}(\eta_b).$ Therefore, the
solution of this maximization is heavily dependent on the efficiency
function given in \eqref{eq34}. While the coding gain can be assumed
to be constant in the limit of very large SIRs, the dependence of
$G$ on $\gamma$ and $b$ is important for our optimization problem.
Since there are no closed-form expressions for $G_b(\gamma)$, we can
use the BER curves available in the literature (for example in
\cite{GoldsmithChua98}) to estimate the coding gain as a function of
$\gamma$ for different modulation levels. Then, using these discrete
values, we can approximate the shape of $G_b(\gamma)$ for different
values of $b$. We have found that the following function gives us a
reasonable estimate for $G$:
\begin{equation}\label{eq39}
    \hat{G}_b(\gamma)= A_b + C_b \tan^{-1} \left( \frac{\gamma
    -\bar{\gamma}_b}{D_b}\right) ,
\end{equation}
where $A_b$, $C_b$, $D_b$, and $\bar{\gamma}_b$ are constants that
only depend on the modulation level and can be determined by trial
and error. The function in \eqref{eq39} is plotted in
Fig.~\ref{fig4} for different values of $b$ for an 8-state
convolutional encoder with rate 2/3. The piece-wise linear curves
obtained from the BER plots are also shown.

Based on the analysis that was done for uncoded systems, the
best-response strategy of a user in a coded system and the
achieved utility at Nash equilibrium depend on $\gamma_b^{*(c)}$,
which is the solution of $f_b^{(c)}(\gamma)= \gamma
f_b^{'^{(c)}}(\gamma)$, and on $\hat{\gamma}_b^{(c)} = f_b^{(c)
-1} (\eta_b)$. We use the approximation given by \eqref{eq39} for
the coding gain in the expression for the efficiency function
given by \eqref{eq34}. While obtaining closed-form expressions for
$\gamma_b^{*(c)}$ and $\hat{\gamma}_b^{(c)}$ is difficult, we can
compute them numerically. Table~\ref{table coded uncoded} compares
the optimum SIRs and the corresponding packet success
probabilities and utilities of the uncoded and coded systems for
different values of $b$ and with no delay
constraints.\footnote{Optimum here refers to the best-response
strategy (i.e., the most energy-efficient solution).} It is seen
from the table that the target SIR (i.e., $\gamma^*$) is lower in
a coded system. Also, comparing the fourth and seventh columns of
Table~\ref{table coded uncoded}, we see that coding improves
user's utility (i.e., energy efficiency).
\begin{table*}[t]
\begin{center} \caption{Comparison between coded and uncoded systems}\label{table coded uncoded}
\begin{tabular}{|c|c|c|c|c|c|c|c|c|}
  \hline
  $b$ & $\gamma_b^*$(dB) & $f_b(\gamma_b^*)$ & $b f_b(\gamma_b^*)/ \gamma_b^*$ & $\gamma_b^{*(c)}$(dB) & $f_b^{(c)}(\gamma_b^{*(c)})$ & $b f_b^{(c)}(\gamma_b^{*(c)})/ \gamma_b^{*(c)}$ \\
  \hline
  \hline
  2 &  9.1 & 0.801  & 0.1978   &  8.1 & 0.947  &  0.2914\\
  4 &  15.7 & 0.785  & 0.0846   &  14.2 & 0.898  &0.1357  \\
  6 &  21.6 & 0.771  & 0.0322  &  20.4 & 0.872  &  0.0448\\
  8 &  27.3 & 0.757  & 0.0112  &  26.3 & 0.847  &  0.160\\
  10&  33.0 & 0.743  & 0.0037   &  31.9 & 0.788  & 0.0051 \\
  \hline
\end{tabular}
\end{center}
\end{table*}

\section{Numerical Results}\label{numericalresults}

\begin{figure}
\begin{center}
\leavevmode \hbox{\epsfysize=3.5in \epsfxsize=3.5in
\epsffile{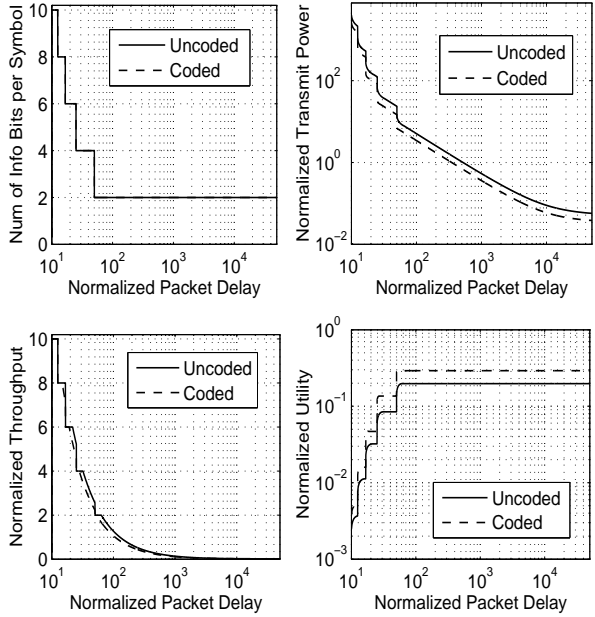}}
\end{center} \caption{Optimum modulation level, transmit power, throughput,
and utility as a function of (normalized) packet delay.}
\label{fig5}
\end{figure}

In this section, we quantify the effect of constellation size on
energy efficiency of a user with a delay QoS constraint. The
packet size is assumed to be 100 bits, and the source rate (in
bps) for the user is assumed to be equal to $0.01B$ where $B$ is
the system bandwidth. We further assume that a user chooses its
constellation size, symbol rate, and transmit power according to
its best-response strategy corresponding to the Pareto-dominant
Nash equilibrium (see Section~\ref{NE}). For the coded system, we
assume an 8-state convolutional encoder with rate 2/3. The code
rate for QPSK is chosen to be 1/2. Fig.~\ref{fig5} shows the
optimum constellation size, transmit power, throughput, and user's
utility as a function of the delay constraint for both uncoded and
coded systems. The results for the coded case are obtained by
using \eqref{eq39} as an approximation for the coding gain. For
all four plots, the packet delay is normalized by the inverse of
the system bandwidth. To keep the spectral efficiency of the two
systems the same, we assume that the number of information bits
transmitted per symbol is the same for both uncoded and coded
systems. The throughput corresponds to the transmission rate for
the user which is obtained by multiplying the symbol rate by the
number of (information) bits per symbol (i.e., $b$), and is
normalized by the system bandwidth. The transmit power and user's
utility are also normalized by $\hat{h}$ and $B\hat{h}$,
respectively.

\begin{figure}
\begin{center}
 \leavevmode \hbox{\epsfysize=3.2in \epsfxsize=3.5in
\epsffile{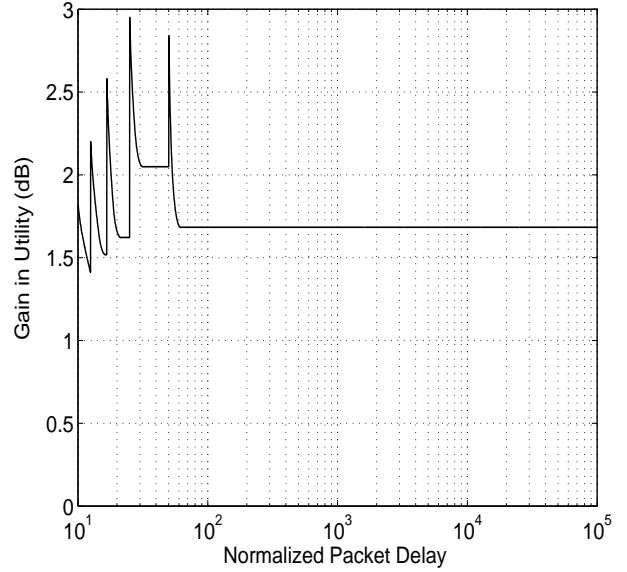}}
\end{center} \caption{Gain in utility due to TCM as a function of (normalized) packet delay.}
\label{fig6}
\end{figure}

Let us for now focus on the uncoded system. When the delay
constraint is large, QPSK (which is the most energy efficient
M-QAM modulation) can accommodate the delay requirement and hence
is chosen by the user. As the delay constraint becomes tighter,
the user increases its symbol rate and also raises the transmit
power to keep the output SIR at 9.1~dB (recall that
$\gamma_b^*=9.1$~dB when $b=2$). In this case, the user's utility
stays constant. Once the symbol rate becomes equal to the system
bandwidth, the user cannot increase it anymore. Hence, as the
delay constraint becomes more stringent, the user is forced to aim
for a higher target SIR to meet its delay requirement. In this
case, the transmission rate stays constant and the transmit power
increases. Hence, the user's utility decreases. As the delay
requirement becomes tighter, a point is reached where the spectral
efficiency of QPSK is not enough to accommodate the delay
constraint. In this case, the user jumps to a higher-order
modulation (i.e., 16-QAM) and the process repeats itself. The
trends are similar for the coded system except that, due to coding
gain, the required transmit power is smaller for the coded system.
This results in an increase in the user's utility. This means
that, for the same number of information bits transmitted per
symbol, the energy efficiency is higher when TCM is used. In
addition, in Fig.~\ref{fig6}, we have plotted the utility gain
achieved due to TCM as a function of normalized packet delay. It
is seen that the gain in energy efficiency due to TCM depends on
the delay constraint and fluctuates between 1.5~dB and 3~dB. In
general, for the same delay constraint, the uncoded system has to
transmit at a slightly higher symbol rate as compared to the coded
system. This is because $f_b^{*(c)} > f_b^{*}$ (see
Table~\ref{table coded uncoded}). As a result,
$\Omega_b^{*(c)}<\Omega_b^*$. The spikes in Fig.~\ref{fig6}
correspond to the cases in which the uncoded system is
transmitting at the maximum possible symbol rate and has to
increase its target SIR to meet the delay constraint. This results
in a drop in the utility of the uncoded system. The coded system
may still be able to meet the delay constraint without increasing
the target SIR.

\begin{figure}
\begin{center}
 \leavevmode \hbox{\epsfysize=3.0in \epsfxsize=3.5in
\epsffile{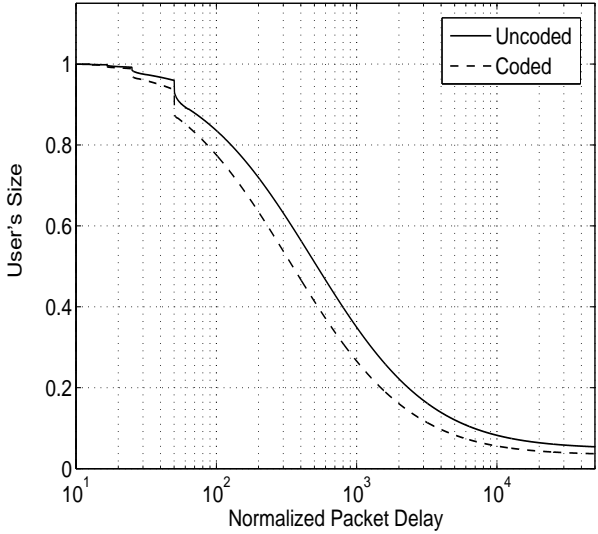}}
\end{center} \caption{User's ``size", $\Phi$, corresponding to the Pareto-dominant Nash equilibrium as a function of (normalized) packet delay.}
\label{fig7}
\end{figure}

Fig.~\ref{fig7} shows the user's ``size", $\Phi$, corresponding to
the Pareto-dominant Nash equilibrium as a function of the packet
delay for both uncoded and coded systems. As explained in
Section~\ref{NE}, $\Phi$ increases as the delay constraint becomes
tighter. This makes sense because a user would need to consume
more network resources to satisfy a more stringent delay. It is
also seen that coding reduces the user's ``size" and, hence,
increases the network capacity. This is because for the same
constellation size, the symbol rate and the SIR are smaller for
the coded system.

\begin{figure}
\begin{center}
\leavevmode \hbox{\epsfysize=3.0in \epsfxsize=3.5in
\epsffile{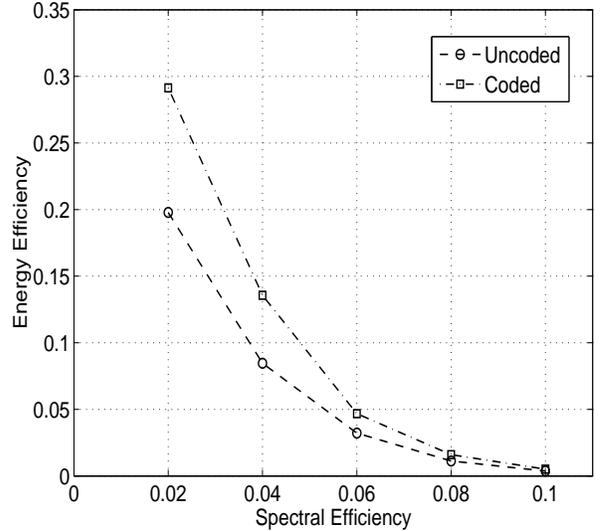}}
\end{center} \caption{Tradeoff between spectral efficiency and energy efficiency.}
\label{fig8}
\end{figure}

We have seen throughout this paper that the strategy that
maximizes the user's energy efficiency is not spectrally
efficient. To illustrate the energy efficiency-spectral efficiency
tradeoff, let us fix the symbol rate to be $0.01B$. For a fixed
constellation size, the user's utility (energy efficiency) is
proportional to $bf(\gamma_b^*)/\gamma_b^*$. The spectral
efficiency is given by $b R_s/B$. By varying the constellation
size, we can quantify the tradeoff between energy efficiency  and
spectral efficiency. This tradeoff is shown in Fig.~\ref{fig8} for
both uncoded and coded systems. In this figure, we have plotted
$bf(\gamma_b^*)/\gamma_b^*$ vs. $b R_s/B$. Different points on the
plot correspond to different values of $b$. The energy
efficiency-spectral efficiency tradeoff is definitely an
interesting and important topic that requires more in-depth
analysis.

\section{Conclusions}\label{conclusion}

We have studied the effects of modulation order on energy
efficiency of wireless networks using a game-theoretic framework.
Focusing on M-QAM modulation, we have proposed a non-cooperative
game in which each user chooses its strategy in order to maximize
its utility while satisfying its delay QoS constraint. The actions
open to the users are the choice of the transmit power, transmit
symbol rate and constellation size. The utility function measures
the number of reliable bits transmitted per joule of energy
consumed and is particularly suitable for energy-constrained
networks. The best-response strategies and the Nash equilibrium
solution for the proposed game have been derived. We have shown
that to maximize its utility (i.e., energy efficiency), the user
must choose the lowest modulation level that can accommodate the
user's delay constraint. Using our non-cooperative game-theoretic
framework, the tradeoffs among energy efficiency, delay,
throughput and constellation size have also been studied and
quantified. In addition, we have included the effects of TCM and
have shown that, as expected, coding increases energy efficiency.
The tradeoff between energy efficiency and spectral efficiency has
also been illustrated.

\end{document}